\documentclass[12pt]{article}

\usepackage{graphicx}
\usepackage{setspace}

\newcommand{\ee}{\end{equation}}
\newcommand{\be}{\begin{equation}}

\newcommand{\dt}{\delta_{t}}
\newcommand{\dx}{\delta_{x}}
\newcommand{\dy}{\delta_{y}}

\newcommand{\pars}{\left(}
\newcommand{\pard}{\right)}

\begin{document}
\title{Analytical calculation of slip flow
in lattice Boltzmann models with kinetic boundary conditions}
\author{{M. Sbragaglia $^{1}$ and  S. Succi$^{2}$}\\ 
{\small $^{1}$ Dipartimento di Fisica and INFN, Universit\`a "Tor Vergata",}\\ 
{\small Via della Ricerca Scientifica 1, I-00133 Roma, Italy}\\
{\small $^{2}$ Istituto per le Applicazioni del Calcolo, CNR,}\\ 
{\small Viale del Policlinico 137, I-00161, Roma, Italy}}

\doublespacing

\maketitle
\begin{abstract}{
We present a mathematical formulation of kinetic boundary
conditions for Lattice Boltzmann schemes in terms of 
reflection, slip, and accommodation coefficients. 
It is analytically and numerically shown that, in the presence of a
non-zero slip coefficient, the Lattice Boltzmann flow develops
a physical slip flow component at the wall.
Moreover, it is shown that the slip coefficient can be tuned 
in such a way to recover quantitative agreement with analytical
and experimental results up to {\it second} order in the Knudsen number.
}
\end{abstract}
\maketitle

\section{Introduction}

Over the last decade, discrete
kinetic methods, and most notably the Lattice Boltzmann (LB), have
known a significant growth for the simulation of a variety 
of complex flows \cite{LBE}.
One of the most valuable properties of the LB method is its flexibility \cite{HSB},       
and easy set up of boundary conditions for complex geometries.
Such a flexibility stems from the fact that the LB dynamics
proceeds along rectilinear trajectories, so that LB
shares the computational and conceptual simplicity of particle methods.
On the other hand, since the LB dynamics typically involves
more dependent variables (discrete distributions) than 
hydrodynamic fields, the mathematical formulation of
boundary conditions is left with some ambiguity, and a systematic
treatment of the subject is still lacking.
As a result, the issue of boundary conditions has become 
one of the most active areas of recent LB research. 
This is especially true for the highly debated question
of the applicability of LB methods to microfluidic 
applications \cite{MICROLBE}.  
As recently shown by Ansumali and Karlin, \cite{AK}, much can
be gained in patterning LB boundary conditions after
the time-honored procedures developed in continuum kinetic 
theory \cite{CER}.
Based on the same philosophy, in this work
we present a general class of homogeneous and isotropic
boundary conditions for lattice Boltzmann models living
on regular lattices. 
This class represents a lattice realization of most popular
boundary conditions in continuum kinetic theory, that is 
diffuse boundary conditions with and without accommodation.
In this work, special emphasis is placed on the issue of slip flow at the
boundary, which plays a central role in microfluidic applications
\cite{MICRO}. Analytical expressions for the slip flow are derived 
for a broad class of boundary conditions and compared
with numerical simulations. Excellent agreement between numerical and analytical results is found over a wide range of parameters.

\section{Formulation of the boundary condition problem}

For the sake of concreteness, we shall refer to the
two-dimensional nine-speed D2Q9 model \cite{LBGK}, 
although the proposed analysis can be extended 
in full generality to any other discrete-speed model 
living on a regular lattice
(for example, the three-dimensional D3Q19 scheme shall 
be used for comparison with numerical details).\\ 
We begin by considering the lattice Boltzmann equation in the
following form: 
 \be
f_{i}(\vec{x}+\vec{c}_{i}\dt,t+\dt)-f_{i}(\vec{x},t)=-\frac{1}{\tau}(f_{i}(\vec{x},t)-f^{(eq)}_{i}(\rho,\vec{u}))+\frac{\dx}{c^{2}} F g_{i}
\ee
where $\vec{c}_{i}$  ($i=0,1,...,8$) is a discrete set of 
velocities: 
$$
\vec{c}_{\alpha}=\left \{ \begin{array}{c c c}
\vec{c}_{0}                                      & = & (0,0)c \\
\vec{c}_{1},\vec{c}_{2},\vec{c}_{3},\vec{c}_{4}  & = & (1,0)c,(0,1)c,(-1,0)c,(0,-1)c \\
\vec{c}_{5},\vec{c}_{6},\vec{c}_{7},\vec{c}_{8}  & = & (1,1)c,(-1,1)c,(-1,-1)c,(1,-1)c. \end{array} \right.
$$
The external source must inject zero mass and $\rho F$ units of momentum
per unit volume and time.
This results in the following constraints on the
forcing coefficients $g_{i}$: 
\[
\sum_i g_i \vec{c}_i = 0 \hspace{.2in} \sum_i g_i \vec{c}_i \cdot \vec{c}_i = 1.
\]
These constraints can be satisfied by choosing
$g_1=-g_3$ and $g_5=g_8=-g_7=-g_6$, thus leaving
us with one free parameter, say $g_5$.\\
We wish to emphasize that since the forcing term is associated with
external forces, the 
$\frac{\vec {F}}{m} \cdot \frac{\partial f}{\partial \vec {v}}$ 
operator in the Boltzmann equation, there is no reason why $g_5$ should take 
the same value in the bulk and at the boundary.
As a result, $g_5$ can be treated as an additional
degree of freedom to describe fluid-wall interactions.\\
The amplitude of the hydrodynamic forcing is chosen in such a way
that, if the boundary velocity is zero, it reproduces a Poiseuille flow with centerline speed $u_0$:
\be\label{force}
F = \frac{8 \nu u_0}{H^2}
\ee
where $H=N_y \delta_y$ is the channel height.
Discrete space and time increments are $\dx=\dy,\dt$, with 
$c=\frac{\dx}{\dt}$, and the equilibrium distribution 
is given by:
\be
f^{(eq)}_{i}(\rho,\vec{u})=w_{i} \rho [1+\frac{(\vec{c_{i}} \cdot \vec{u})}{c_s^{2}}+\frac{1}{2}\frac{(\vec{c_{i}} \cdot 
\vec{u})^{2}}{c_s^{4}}-\frac{1}{2}\frac{u^{2}}{c_s^{2}}]
\ee
with $w_{0}=4/9$, $w_{1,2,3,4}=1/9$, $w_{5,6,7,8}=1/36$ and
$c_s^2=\sum_i w_i c_i^2=c^2/3$ the lattice sound speed. 
We consider a system whose $y$-coordinates lie between $[0,N_{y}\dy]$, $N_{y}$ 
being the number of grid points with solid walls located at  
$-\frac{1}{2}\dy$ (south wall), and $N_{y}\dy+\frac{1}{2}\dy$ (north wall). 
In the following, we will refer to the north wall in order to study 
the boundary condition problem.
A general way of formulating the boundary conditions is:
\be
f_{j}(\vec{y})=\sum_i {\cal K}_{j,i}(\vec{x},\vec{y},t)f_{i}(\vec{x})
\ee
where the matrix ${\cal K}_{j,i}$ is the discrete analogue 
of the boundary scattering kernel expressing the fluid-wall
interactions.
In the above $\vec{y}=\vec{x}+\vec{c}_i$  
runs over the surface of the wall and the indices $i,j$ 
stand for incoming and outgoing velocities respectively
($j=4,7,8$ and $i=2,5,6$ for the specific case of  D2Q9)).  
To guarantee conservation of mass and normal momentum, the
following sum-rule applies:
\be\label{norm}
\sum_{j} {\cal{K}}_{j,i}=1
\ee
and upon the assumption of stationarity  of the fluid-wall
interaction, we can drop the dependence on $t$ and write:
\be
f_{j}=\sum_i {\cal K}_{j,i}f_{i}.
\ee
We now focus our attention on the isotropic homogeneous case. 
The most general form for ${\cal K}_{j,i}$ is:
\be
{\cal K}_{j,i}={\cal K}_{j,i}(|\hat{c}_{i} \cdot \hat{c}_{j}|,|\hat{c}_{i} \cdot \hat{n}|)
\ee 
where $\hat{n}$ is the outward unit normal to the surface boundary.
The dependence on the second argument is necessary to develop a 
non-zero slip component in the stream-wise direction.\\
For the case of D2Q9  (same goes for D3Q19), the quantity 
$|\hat{c}_{i} \cdot \hat{n}|$  can take only two values, related to 
the two only possible angles that can be formed between a 
generic incoming velocity and the normal $\hat{n}$, in our case 
$\alpha_{1}=\frac{3}{4} \pi$ and $\alpha_{2}=\pi$. 
Explicitly,
\be
\left( \begin{array}{c}
        f_{7}(x,N_{y}\dy+\dy) \\
        f_{4}(x,N_{y}\dy+\dy) \\
        f_{8}(x,N_{y}\dy+\dy)   \end{array} \right)={\cal K} \left( \begin{array}{c}
        f_{5}(x-\dx,N_{y}\dy) \\
        f_{2}(x,N_{y}\dy) \\
        f_{6}(x+\dx,N_{y}\dy)   \end{array} \right)
\ee
where
\be
{\cal K} =  \left( \begin{array}{c c c}
            {\cal K}_{\pi,\alpha_{1}} & {\cal K}_{\frac{3}{4}\pi,\alpha_{1}} & {\cal K}_{\frac{\pi}{2},\alpha_{1}}  \\ 
            {\cal K}_{\frac{\pi}{2},\alpha_{2}} & {\cal K}_{\pi,\alpha_{2}} & {\cal K}_{\frac{\pi}{2},\alpha_{2}} \\ 
            {\cal K}_{\frac{\pi}{2},\alpha_{1}} & {\cal K}_{\frac{3}{4}\pi,\alpha_{1}} & {\cal K}_{\pi,\alpha_{1}}  \end{array} \right).
\ee
Under the assumption of Stokes-flow, it can be shown that 
in the limit of zero spacings, the above scattering kernel
provides an analytical expression for the slip velocity that, 
in the continuum limit of  small time and space increments, reads:
\be\label{slip}
u_{slip}= A_{{\cal K}} Kn \left|\frac{\partial u_x}{\partial \hat{n}}\right|_{wall}   + B_{{\cal K}} Kn^2  \left|\frac{\partial^2 u_x}{\partial \hat{n}^2}\right|_{wall}\ee
where 
$$
A_{{\cal K}}=\pars \frac{1-{\cal K}_{\pi,\alpha_{1}}+{\cal K}_{\frac{\pi}{2},\alpha_{1}}}{1+{\cal K}_{\pi,\alpha_{1}}-{\cal K}_{\frac{\pi}{2},\alpha_{1}}} \pard  \frac{c}{c_s}
$$
$$
B_{{\cal K}}=\frac{12 g_1}{(1+{\cal K}_{\pi,\alpha_{1}}-{\cal K}_{\frac{\pi}{2},\alpha_{1}})} + \pars \frac{1-{\cal K}_{\pi,\alpha_{1}}+{\cal K}_{\frac{\pi}{2},\alpha_{1}}}{1+{\cal K}_{\pi,\alpha_{1}}-{\cal K}_{\frac{\pi}{2},\alpha_{1}}} \pard \pars 12 g_5 -\frac{c^2}{c^2_s} \pard.
$$
In the above, $\tau_{f} = \frac{\tau}{\dt}$ is the relaxation time 
of the continuum Boltzmann equation (proportional to the Knudsen number) and $\hat{n}$ is the inward normal to the wall. 
The case of finite spacings can be treated exactly (see Appendix), but for 
sake of simplicity, here we report only the continuum case.\\ 
We  wish to emphasize that what we call here 'slip velocity' 
is the velocity at the nodes nearest to the wall 
($(x,-\frac{1}{2}\dy)$ and $(x,N_{y}\dy+\frac{1}{2}\dy)$); 
to compute the velocity at the wall, an extrapolation 
of the parabolic profile at $(x,-\frac{1}{2}\dy)$ is required.\\
Our analytical formula provides the slip velocity 
dependence on the stresses at the wall, on the Knudsen number and   on the matrix elements of 
the scattering kernel, as well as on the forcing
weights along the tangential and diagonal directions,
$g_1$ and $g_5$ respectively.
We now proceed to a more direct physical
interpretation in terms of reflection and 
accommodation coefficients. 

\subsection{Slip-Reflection model (SR)}

The first example involves two parameters $r,s$, representing 
the probability for a particle to be bounced back 
and slipped forward, respectively.
The boundary kernel takes the form (see \cite{SLIP}):
\be
{\cal{K}} =  \left( \begin{array}{c c c}
r & 0 & s  \\ 
0 & r+s & 0 \\ 
s & 0 & r  \end{array} \right).
\ee
Obviously, the two parameters are not independent and
must be chosen such that $r+s=1$.
In this case the slip velocity (\ref{slip}) reads :
\be
\label{usliprs}
u_{slip}= A Kn  \left|\frac{\partial u_x}{\partial \hat{n}}\right|_{wall}  + B Kn^2 \left|\frac{\partial^2 u_x}{\partial \hat{n}^2}\right|_{wall}
\ee
where
\be
\label{AB}
A = \frac{c}{c_s}\frac{1-r}{r}  \hspace{.2in} 
B = \frac{c^2}{c_s^2} (1- 4 g_5). 
\ee
From the first expression, it is clear that the coefficient
$A$ is equivalent to Maxwell's first order slip velocity
$$u_{slip}= \frac{2-\sigma}{\sigma} Kn \left|\frac{\partial u_x}{\partial \hat{n}}\right|_{wall}  $$
with $\sigma=2r$ \cite{MAX} the well 
known Maxwell accommodation coefficient.
It should be noted that in the limit of pure bounce-back
($r=1$) the leading term disappears, and one is left with
a purely quadratic dependence on the Knudsen number.
This quadratic term stems from the tangential populations
$f_1$ and $f_3$; the independence of the
coefficient $B$ of $r$ is due to the fact that the
tangential populations are evolved according to the same
LB dynamics as in the bulk. 
We also note that in the limit of pure slip $r \rightarrow 0$, 
the slip flow tends to diverge, as it must be since this limit
corresponds to zero friction at the wall. 
This contrasts with previous results \cite{SLIP}, which reported 
a finite slip length even at vanishingly small values of $r$.
The explanation is that those results were not converged in time.\\
It is instructive to compare (\ref{usliprs}) with the
corresponding analytical solution for the fully
continuum case (also with a continuum velocity phase space), namely \cite{CER}
\[
u_{slip} = 1.146 Kn \left|\frac{\partial u_x}{\partial \hat{n}}\right|_{wall} + 0.907 Kn^2 \left|\frac{\partial^2 u_x}{\partial \hat{n}^2}\right|_{wall}.
\]
First, we notice that there exists a value of $r$ such that
the leading term can be reproduced exactly, that is:
$r^* \sim 0.603$.  
The quadratic term can also be reproduced exactly by choosing
$g_5=0.175$ and $g_1=0.15$.
Note that this choice does not affect the bulk behaviour 
where the Stokes equation is still valid.\\
Experimental data on slip flow are sometimes interpreted 
by assuming a Stokes flow   in the bulk, coupled to a second
order boundary condition of the form (\ref{usliprs}).
In fact, in this way,  upon the integration of the Stokes equation
with the second order boundary slip it can be shown that the mass flow rate $Q_{c}$ of the channel is given by:
\be\label{slipcoeff}
Q_{c}=S Q_{p}
\ee
with $Q_{p}$ the Poiseuille mass flow rate and $S$ a dimensionless number called slip coefficient depending on the Knudsen number :
$$
S=1+6 A Kn+12 B Kn^{2}.
$$
In recent experimental works \cite{TAB}
the slip coefficient has been calculated for Helium and Nitrogen
and a good experimental fit has been obtained with a second order 
slip at the boundaries that gives $A=1.2 \pm 0.05$, $B=0.23 \pm 0.1$ for Helium, and
$A=1.3 \pm 0.05$, $B=0.26 \pm 0.1$ for Nitrogen.
It is interesting to note
that both sets of coefficients can be exactly reproduced
by our model by choosing
$r \sim 0.59 , g_5 \sim 0.22,g_1 \sim 0.06$ and
$r \sim 0.59 , g_5 \sim 0.23,g_1 \sim 0.04$ for Helium and Nitrogen respectively.
More generally, being subject to the constraint $g_5 \le 1/4$,
our model permits to reproduce second order slip coefficients in the
range $0 \leq B \leq 3$.

\subsection{Slip-Reflection-Accommodation model (SRA)}

The second example is a straightforward generalization of 
the previous model, which 
is characterized by a third parameter, $a$, related to the presence of 
wall-relaxation (accommodation) phenomena at the fluid-solid interface.
By accommodation we imply that the energy of the incoming and outgoing
particles is not the same because the outgoing 
ones are re-injected into the bulk with equilibrium weights. 
The SRA boundary kernel reads:
\be
{\cal{K}} = \left( \begin{array}{c c c}
r + a W_{2} & a W_{2} & s+ a W_{2}  \\ 
a W_{1} & r+s+a W_{1} & a W_{1} \\ 
s+ a W_{2} & a W_{2} & r+ a W_{2}  \end{array} \right)
\ee
with the normalization constraints:
$$r+s+a=1$$ and 
$W_{2}=1/6, W_{1}=2/3 $.
The presence of the weights $W_1$ and $W_2$ is due to the fact that they
are the discrete analogue of the perfect accommodation kernel, i.e.
a uniform Maxwell distribution at wall temperature.
The SRA slip velocity has exactly the same form as (\ref{AB})
with the plain replacement 
$$r'=r+a/2.$$
As expected, this implies that the accommodation coefficient
$a$ results in a smaller slip flow and
this shows that the SRA model is basically equivalent to the SR one.\\
It is known from continuum kinetic theory that a single
accommodation coefficient is not sufficient for the
quantitative interpretation of experimental data.
To cope with this problem, nearly four decades ago
Cercignani and Lampis proposed a generalization of Maxwell's
diffuse-boundary model which includes two accommodation coefficients,
normal and tangential to the walls \cite{CL}.
The lattice analogue of the Cercignani-Lampis kernel 
is readily computed.
However, the analysis shows that only the accommodation
along the tangential direction plays a role, while the
one along the normal direction disappears. 
This is a pathology of all lattice models where incoming and 
outgoing normal velocities have the same magnitude. 
Therefore, we shall not consider this model any further in this work but
the use of the lattice Cercignani-Lampis kernel 
might bear an interest for the case of thermal LB schemes.

\section{Numerical validation}

\begin{figure}[t!]
\begin{center}
\includegraphics[scale=1.0]{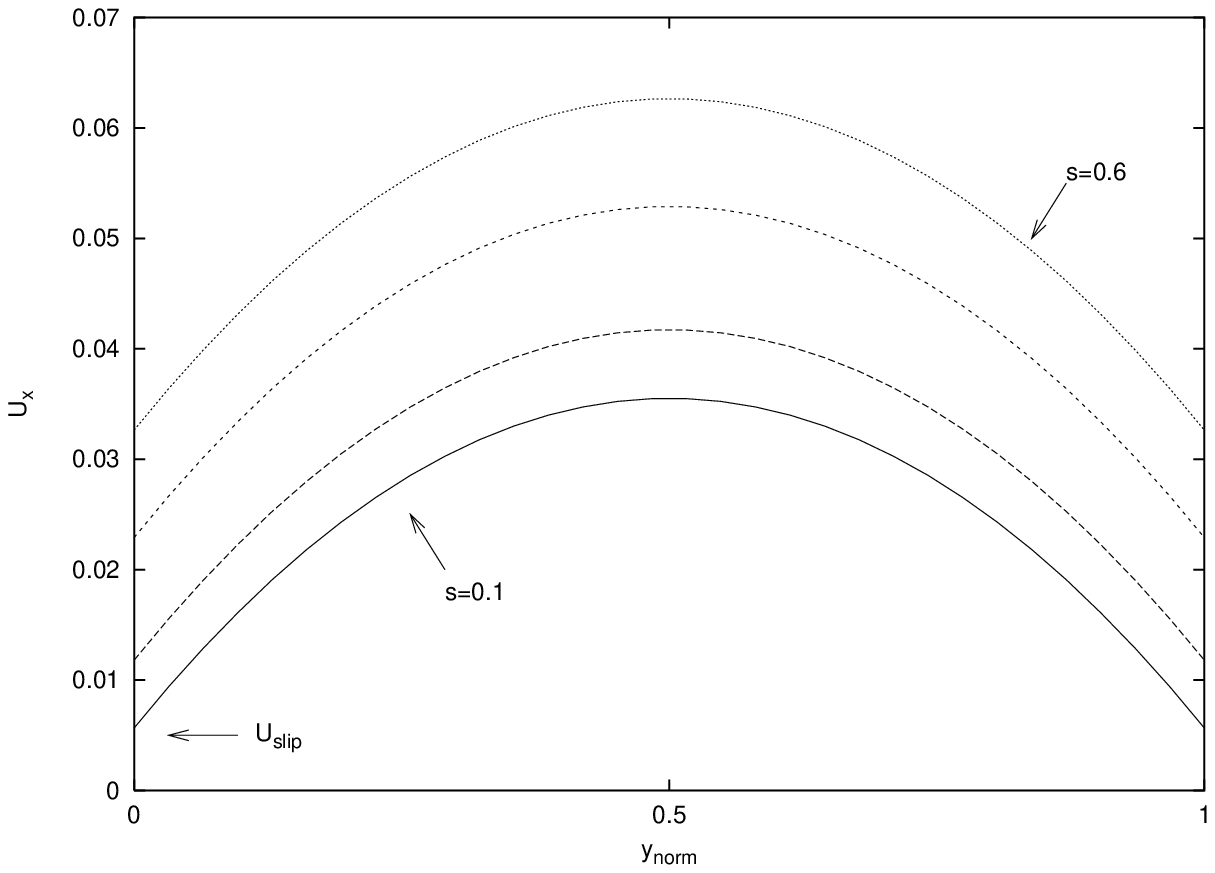}
\end{center}
\caption{}
\label{SR}
\end{figure}

We now present a numerical study aimed at validating
our analytical results for the $SR$ and $SRA$ kernels. 
To this purpose, we have
performed numerical simulations of a channel flow
with the $3d$ lattice Boltzmann model  with $19$ discrete  speeds (D3Q19). 
First, we consider the case of equi-balanced 
external forcing among all biased populations 
(the equivalent of $g_5=g_1$ in the case of $D2Q9$). 
The channel has dimensions $N_{x}=64,N_{y}=32,N_{z}=32$, the  Knudsen 
number is $Kn=0.08$ and the forcing has been fixed 
so as to reproduce a center-channel velocity  $u_{0}=0.03$
in the limit of a Poiseuille flow (\ref{force}).\\
In Figure \ref{SR} we show the center-channel profile in the 
stream-wise direction for the $RS$ case. 
By varying the free parameter $r$ between $0.1$ and $1$
the profile ``shifts'' with no changes in its 
concavity, which is fixed by the external forcing term $F$ 
and the viscosity $\nu$, both kept constant in the simulations.\\
In Figure \ref{SRslip} we present a comparison between the slip velocity, as 
extracted from numerical data, and our analytical 
results for the {\it discrete} case (see Eq. (\ref{analisi}) in the
Appendix). 
An excellent agreement between numerical and analytical results
in the range $0.1<r<1$ is clearly observed. For sake of completeness we have also compared the mass flow rate normalized to the pure bounce back case with our analytical solution (inset of Figure \ref{SRslip}) and, as expected, also here an excellent agreement is found.\\
In Figure \ref{SRA}, we show the effect of the accommodation parameter $a$  
in reducing the slip flow.
The parameter is chosen as $a=0.3$ and the values of $r,s$ are chosen accordingly. The slip velocity  is the same 
as the case $SR$, only with a renormalized reflection coefficient $r^{'}=r+a/2$ as can be clearly seen in the inset.\\
We have also performed a set of numerical simulations with different repartitions of the external forcing term, in order to study the slip properties as a function of the Knudsen number. 
In the case of $SR$ kernels, fixing the bounce-back parameter 
to $r=0.59$ and  the center-channel velocity to $u_0=0.01$
in the limit of Poiseuille flow, we have studied 
the slip velocity as a function of the Knudsen number 
$Kn$ in the range $0<Kn<0.8$. 
We have chosen an unbalanced bipartition with a forcing coefficient equal to $0.23$ for the incoming populations
 in order to reproduce (see Figure \ref{SRtab}) the experimental results showed in \cite{TAB} for the case of Helium; the numerical results confirm our analysis and are indistinguishable from experimental data 
up to second order terms in the  Knudsen number.

\section{Conclusions}

Summarizing, we have presented a unified formulation
for a broad class of homogeneous and isotropic boundary conditions for 
lattice Boltzmann models living on regular lattices. 
Analytical expressions for the slip flow at the fluid-solid
boundary have been derived and successfully compared with
numerical simulations of three-dimensional channel flow.
The main conclusion is as follows: 
by allowing a non-zero slip coefficient, the 
Lattice Boltzmann flow develops a slip flow component which 
can be matched exactly to analytical and experimental
data up to second order in the Knudsen term, well
inside the transition regime ($0<Kn<0.8$).
This means that the lattice Boltzmann scheme with kinetic 
boundary conditions {\it can} be used to predict slip flow at 
finite Knudsen numbers well beyond the strict hydrodynamic limit.
Of course, the extension of the present model to
more realistic situations, involving complex geometries and/or
inhomogeneities \cite{TRO,YEO} and thermal
effects \cite{GARCIA2} remains an open issue.

\clearpage

\section*{Acknowledgments}

We are grateful to S. Ansumali, R.Benzi, I. Karlin and F.Toschi for useful discussions. 
We thank L. Biferale for a careful and critical reading of the manuscript. 

\section{Appendix}

We impose stationary condition on the node $(x,N_{y}\dy)$ and,
under the assumption of homogeneity, we drop the $x$ dependence and write :
\be
\left \{  \begin{array}{l}
            {\cal{K}}_{\pi,\alpha_{1}}f_{5}(N_{y}\dy)+{\cal{K}}_{\frac{3}{4} \pi,\alpha_{1}} f_{2}(N_{y}\dy)+ {\cal{K}}_{\frac{\pi}{2},\alpha_{1}} f_{6}(N_{y}\dy)-\frac{\dx}{c^{2}}Fg_5=f_{7}(N_{y}\dy) \\
            {\cal{K}}_{\frac{\pi}{2},\alpha_{1}} f_{5}(N_{y}\dy)+ {\cal{K}}_{\frac{3}{4} \pi,\alpha_{1}}   f_{2}(N_{y}\dy)+ {\cal{K}}_{\pi,\alpha_{1}}  f_{6}(N_{y}\dy)+\frac{\dx}{c^{2}}Fg_5=f_{8}(N_{y}\dy)\\
            (f_{1}-f_{3})(N_{y}\dy)-(f^{(eq)}_{1}(\rho,\vec{u})-f^{(eq)}_{3}(\rho,\vec{u}))=2F\tau \frac {\dx}{c^{2}}g_1 . \end{array} \right.
\ee
We can now write for the velocity in the $x$-direction at the height $(N_{y}\dy)$:
\be\label{impulse}
\frac{1}{3} \rho u_{slip}=(2g_1\tau+2g_5)F\frac{\dx}{c}+c( 1-{\cal{K}}_{\pi,\alpha_{1}}  +{\cal{K}}_{\frac{\pi}{2},\alpha_{1}}  )\zeta(N_{y}\dy)
\ee
with
\be\label{deff}
f_{5}(x,y)-f_{6}(x,y)=\zeta(y).
\ee
Using the equations for $f_{5}$ and $f_{6}$ in the bulk, and under the 
stationarity assumption, we obtain:
\be
\left \{ \begin{array}{l}
f_{5}(x+c\dt,y+c\dt)-f_{5}(x,y)=-\frac{1}{\tau}(f_{5}(x,y)-f^{(eq)}_{5}(\rho,\vec{u}))+\frac{\dx}{c^{2}} Fg_5\\
f_{6}(x-c\dt,y+c\dt)-f_{6}(x,y)=-\frac{1}{\tau}(f_{6}(x,y)-f^{(eq)}_{6}(\rho,\vec{u}))-\frac{\dx}{c^{2}} Fg_5. \end{array} \right.  
\ee
Dropping the $x$ dependence (homogeneity), upon 
the definition (\ref{deff}), we have
\be
\zeta(y)-\zeta(y-c\dt)=-\frac{1}{\tau}\zeta(y-c\dt)+\frac{\rho u_{x}(y-c\dt)}{6 \tau c} + 2Fg_5\frac{\dx}{c^{2}}
\ee
In the limit $\delta_x,\delta_y,\delta_t \rightarrow 0$ we obtain an $O.D.E.$ which can be solved
exactly . However, by considering finite spacings such that $\delta_x=\delta_y=c=1$, we have:
\be\label{master}
\zeta(j)-\zeta(j-1)=-\frac{1}{\tau}\zeta(j-1)+\frac{\rho u_{x}(j-1)}{6 \tau } + 2Fg_5 \hspace{.4cm} 1 \leq j \leq N_{y}
\ee
where $j$ stands  runs over the channel height. 
This finite difference equation can be solved exactly. 
The solution can be divided in two terms: the homogeneous term $\zeta^{hom}(j)$ 
and a particular term $\zeta^{part}(j)$. For the homogeneous term we have:
\be\label{omo}
\zeta^{hom}(j)=\pars 1-\frac{1}{\tau}\pard ^{j}
\ee
while for the particular term we use the method of variation of constants:
\be\label{part}
\zeta^{part}(j)=\pars 1-\frac{1}{\tau}\pard^{j-1} \sum_{k=0}^{j-1}\pars 1-\frac{1}{\tau} \pard^{-k}[ \frac{\rho u_{x}(k)}{6 \tau } + 2Fg_5].
\ee
Summing the two contributions (\ref{part}) and (\ref{omo}) 
we have an explicit form for the general solution of (\ref{master})
\be
\zeta(j)=C\pars 1-\frac{1}{\tau}\pard ^{j}+\pars 1-\frac{1}{\tau}\pard ^{j-1} \sum_{k=0}^{j-1}\pars 1-\frac{1}{\tau}\pard ^{-k}[ \frac{\rho u_{x}(k)}{6 \tau } + 2Fg_5]
\ee
where $C$ is a constant to be fixed based upon boundary conditions. 
Since $\zeta(N_{y})$ enters in (\ref{impulse}), we have:
\be
\zeta(N_{y})=C\pars1-\frac{1}{\tau}\pard^{N_{y}}+\pars1-\frac{1}{\tau}\pard^{N_{y}-1} \sum_{k=0}^{N_{y}-1}\pars 1-\frac{1}{\tau}\pard^{-k}[ \frac{\rho u_{x}(k)}{6 \tau } + 2Fg_5]
\ee
and under the assumption of low-Knudsen numbers ($N_{y} / \tau >> 1 $), 
$\zeta(N_{y})$ is well approximated by 
$\zeta^{part}(N_{y})$, since $(1-\frac{1}{\tau})^{N_{y}} \rightarrow 0$. 
As a result:
\be
\zeta(N_{y})=\pars 1-\frac{1}{\tau} \pard ^{N_{y}-1} \sum_{k=0}^{N_{y}-1}\pars 1-\frac{1}{\tau}\pard ^{-k}[ \frac{\rho u_{x}(k)}{6 \tau } + 2Fg_5]
\ee
which substituted in (\ref{impulse}), and setting $\rho=1$, returns the
following expression for the velocity at the height ($N_{y}$):
\be
u_{slip}=(6\tau g_1+6g_5)F+\frac{1}{2} ( 1-{\cal{K}}_{\pi,\alpha_{1}}  +{\cal{K}}_{\frac{\pi}{2},\alpha_{1}}  )[\pars 1-\frac{1}{\tau}\pard ^{N_{y}-1} \sum_{k=0}^{N_{y}-1}\pars 1-\frac{1}{\tau}\pard^{-k}(\frac{u_{x}(k)}{\tau } + 12 Fg_5)].
\ee
Next we impose a low-Reynolds regime, by specifying the velocity 
field as a (symmetric) parabolic profile:
\be
\left \{ \begin{array} {l l} 
u_{x}(j)=aj^{2}+bj+u_{slip}  &  0 \leq j \leq N_{y}             \\
b=-a N_{y}            &  {\mbox {\small{since}}} \hspace{.2in}   u_{x}(N_{y})=u_{x}(0)          \\
a=-\frac{ F}{2 \nu}    &  {\mbox {\small{since}}} \hspace{.2in}  \nu \partial_{jj}u_{x}(j)=-F       \end{array} \right.            
\ee
and we obtain for the slip velocity:
\be\label{analisi}
u_{slip} =\frac{(12\tau g_1+12 g_5)F}{({\cal{K}}_{\pi,\alpha_{1}}-{\cal{K}}_{\frac{\pi}{2},\alpha_{1}}+1)}+  \pars \frac{1-{\cal{K}}_{\pi,\alpha_{1}}+{\cal{K}}_{\frac{\pi}{2},\alpha_{1}}}{{\cal{K}}_{\pi,\alpha_{1}}-{\cal{K}}_{\frac{\pi}{2},\alpha_{1}}+1} \pard  [12F\tau g_5 +I_{1}+I_{2}]
\ee
where:
\be
I_{1}=\frac{a}{\tau} \pars 1-\frac{1}{\tau}\pard^{N_{y}-1} \sum_{k=0}^{N_{y}-1}k^{2}\pars 1-\frac{1}{\tau}\pard ^{-k}
\ee
\be
I_{2}=\frac{b}{\tau} \pars 1-\frac{1}{\tau}\pard ^{N_{y}-1} \sum_{k=0}^{N_{y}-1}k\pars 1-\frac{1}{\tau}\pard ^{-k}.
\ee



\newpage

\centerline{FIGURE 1}
Velocity profiles with $SR$ kernels. We plot the center-channel profile in the stream-wise direction as a function of the normalized distance from the wall $y_{norm}=y/H$, being $H$ the channel height. Different values of the parameters $(r,s)$ are considered. From top to bottom we plot the following cases: $(0.4,0.6),(0.5,0.5),(0.7,0.3),(0.9,0.1)$. In all simulations the Knudsen number is  $Kn=0.08$ and the forcing term has been fixed to reproduce a center channel velocity $u_0=0.03$ in the limit of a Poiseuille flow (\ref{force}).\\
\centerline{FIGURE 2}
Slip velocity for the $SR$ kernels. We plot $u_{slip}$ vs $r$ for different values of the bounce back parameter $r$. The data of numerical simulations (boxes) are compared with the analytical estimate (\ref{analisi}) (dotted line). The agreement is excellent all over the range $0.1<r<1$. Inset: numerical data ((boxes) representing  the mass flow rate normalized to its pure bounce-back value ($Q_{norm}$) are compared with our analytical estimate (dotted line).\\
\centerline{FIGURE 3}
Velocity profiles with the $SRA$ kernels. We plot the center channel profile in the stream-wise direction as a function of the normalized distance from the wall $y_{norm}=y/H$, being $H$ the channel height. Different values of the parameters $(a,r,s)$ are considered. We analyze the following plots: $(0.0,0.5,0.5)$ (top plot) and $(0.3,0.5,0.2)$ (bottom plot). Inset: analysis for the slip velocity in the channel in the case $(a,r,s)$ with $a=0.3$. The data of numerical simulations (boxes) are compared with the analytical estimate (\ref{analisi}) (dotted line) with a normalized reflection coefficient $r^{'}=r+a/2$. The agreement is excellent all over the range $0.2<r<0.7$.\\ 
\centerline{FIGURE 4}
Analysis for the slip coefficient (\ref{slipcoeff}) in the channel as a function of the Knudsen number. The data of numerical simulations (circles) with a properly unbalanced repartition of the external forcing are compared with the experimental results 
(triangles) given in \cite{TAB} and with the analytical expression (\ref{slipcoeff}) with $A=1.2$ and $B=0.23$ (dotted line). The linear full accommodation case ($u_{slip}=Kn\left|\frac{\partial u_x}{\partial \hat{n}}\right|_{wall}   $)   is also plotted for a comparison. Inset: the slip velocity as a function of the  Knudsen number is studied with the same repartition of the external forcing as before. We compare the numerical data (circles) with our analytical expression (dotted line) including, as before, terms up to the second order in the Knudsen number. In all the simulations we have used a $SR$ kernel with a bounce back parameter $r=0.59$.\\
\newpage

\begin{figure}[t!]\label{SRslip}
\begin{center}
\includegraphics[scale=1.0]{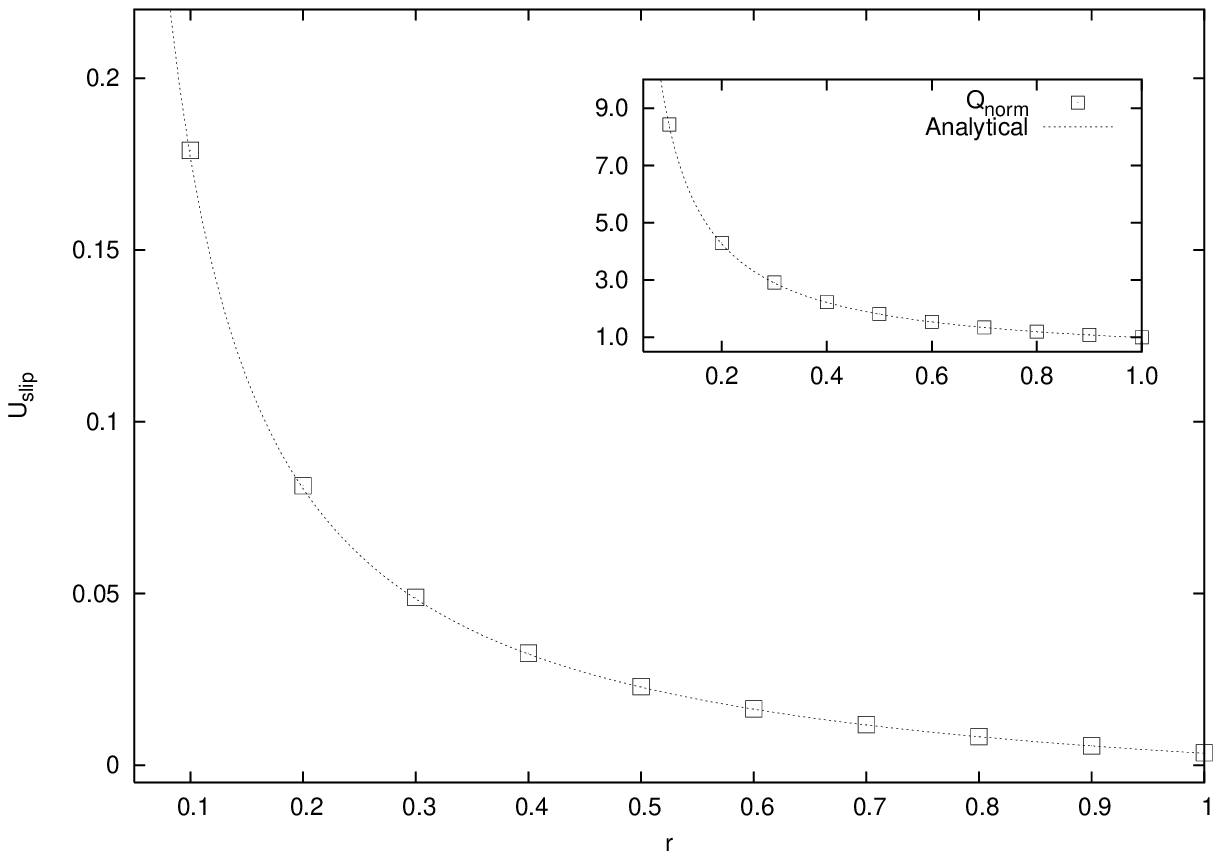}
\end{center}
\caption{}
\label{SRslip}
\end{figure}

\begin{figure}[t!]
\begin{center}
\includegraphics[scale=1.0]{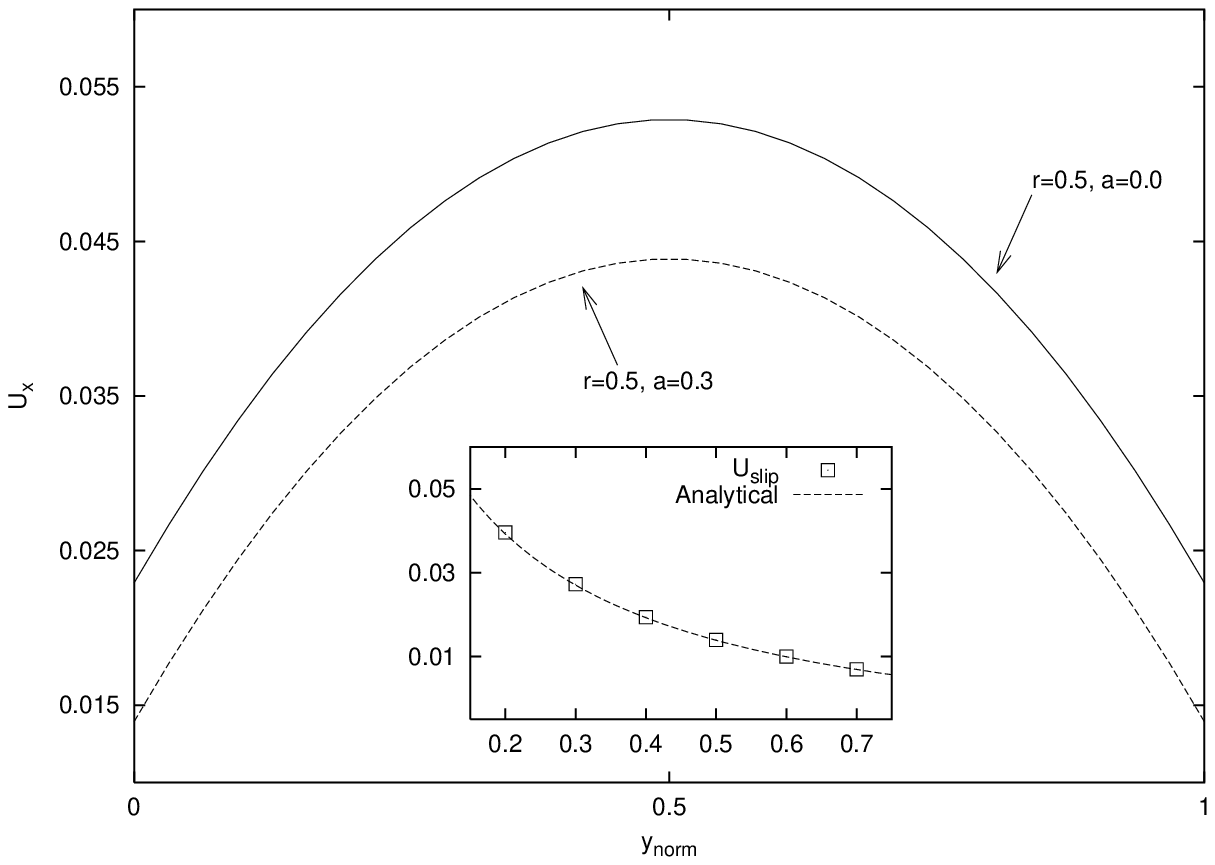}
\end{center}
\caption{}
\label{SRA}
\end{figure}

\begin{figure}[t!]
\begin{center}
\includegraphics[scale=1.0]{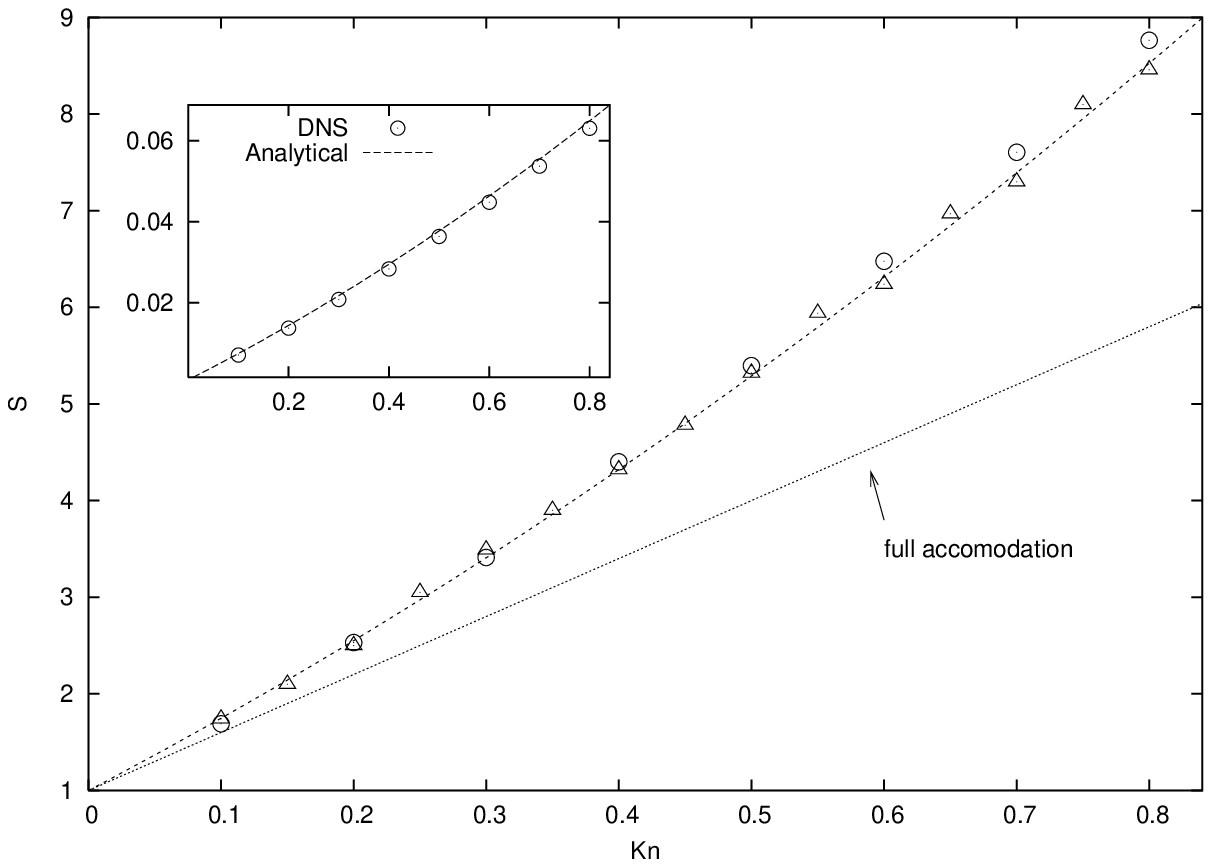}
\end{center}
\caption{}
\label{SRtab}
\end{figure}

\end{document}